\documentclass[reprint,twocolumn,superscriptaddress,secnumarabic,amssymb, nobibnotes, aps, pra]{revtex4-1}

\usepackage{amsmath}    
\usepackage{graphicx}    
\usepackage{verbatim}    
\usepackage{color}           
\usepackage{subfigure}   
\usepackage{hyperref}    
\usepackage{verbatim}  
\usepackage{epstopdf}
\def\laco*{LaCoO$_3$}

\begin{document}

\title{Excitonic dispersion of the intermediate-spin state in LaCoO$_3$ revealed by \\ resonant inelastic X-ray scattering}

\author{Ru-Pan Wang}
\thanks{These two authors contributed equally to this work.}
\affiliation{Debye Institute for Nanomaterials Science, Utrecht University, Universiteitsweg 99, 3584 CG Utrecht, The Netherlands}

\author{Atsushi Hariki}
\thanks{These two authors contributed equally to this work.}
\author{Andrii Sotnikov}
\affiliation{Institute for Solid State Physics, TU Wien, 1040 Vienna, Austria}

\author{Federica Frati}
\affiliation{Debye Institute for Nanomaterials Science, Utrecht University, Universiteitsweg 99, 3584 CG Utrecht, The Netherlands}

\author{Jun Okamoto}
\author{Hsiao-Yu Huang}
\author{Amol Singh}
\author{Di-Jing Huang}
\affiliation{National Synchrotron Radiation Research Center, No.101 Hsin-Ann Road, Hsinchu Science Park, Hsinchu 30076, Taiwan}

\author{Keisuke Tomiyasu}
\affiliation{Department of Physics, Tohoku University, Aoba, Sendai 980-8578, Japan}

\author{Chao-Hung Du}
\affiliation{Department of Physics, Tamkang University No.151 Yingzhuan Rd., Tamsui Dist., New Taipei City 25137, Taiwan}

\author{Jan Kune\v s}
\thanks{kunes@ifp.tuwien.ac.at}
\affiliation{Institute for Solid State Physics, TU Wien, 1040 Vienna, Austria}

\author{Frank M. F. de Groot}
\thanks{F.M.F.deGroot@uu.nl} 
\affiliation{Debye Institute for Nanomaterials Science, Utrecht University, Universiteitsweg 99, 3584 CG Utrecht, The Netherlands}

\date{\today}

\begin{abstract}
We report Co $L_3$-edge resonant inelastic X-ray scattering on LaCoO$_3$ at 20~K. 
We observe excitations with sizable dispersion that we identify as 
intermediate-spin (IS) states. 
Theoretical calculations that treat the IS states as mobile excitons 
propagating on the low-spin (LS) background support the interpretation.  
The present result shows that mobility substantially reduces the
energy of IS excitations in part of the Brillouin zone, which makes them
important players in the low-energy physics of LaCoO$_3$
together with immobile high-spin (HS) excitations.
\end{abstract}

\maketitle

\section{introduction}

More than 50 years ago the electron-hole attraction was
proposed to drive narrow gap semiconductors or semimetals to a new phase, the excitonic insulator. The experimental proof of its existence in bulk materials remains elusive. 
In strongly correlated insulators, the proximity of 
the excitonic insulator phase is reflected by the presence of
dispersive electron-hole excitations 
with a small gap above a singlet ground state~\cite{merchant14}.
Recently, such an excitation spectrum was proposed to be realized in perovskite oxide LaCoO$_3$~\cite{Sotnikov2016sr}.
The purpose of the present study is to test the proposal experimentally using  Co $L_3$-edge resonant inelastic X-ray scattering (RIXS).

At low temperature LaCoO$_3$ is a non-magnetic insulator with Co ions
in the low-spin (LS, $S=0$, $t_{2g}^6e_g^0$) ground state.
Upon heating it undergoes a crossover to a paramagnetic Curie-Weiss insulator ($T\sim100$~K) and,
eventually, a Curie-Weiss metal ($T\sim500$~K)~\cite{Goodenough1958jpcs,
heikes64,Raccah1967pr,Abbate1993prb,Asai1998jpsj,Stolen1997prb,Haverkort2006prl}. 
Traditionally, the spin-state crossover
has been described as a thermal population of excited atomic multiplets.
Despite its long history, the opinion on the nature of the first excited Co$^{3+}$ multiplet remains split between the high-spin (HS, $S=2$, $t_{2g}^4e_g^2$)~\cite{Tanabe1954jpcj,deGroot1990prb,Haverkort2006prl}
 and intermediate-spin (IS, $S=1$, $t_{2g}^5e_g^1$,)~\cite{Korotin1996prb} states. Both scenarios are allegedly supported by experiments,
see, e.g.,~Refs.~\cite{Haverkort2006prl,Podlesnyak2006prl,Noguchi2002prb} and~\cite{Zobel2002prb,Maris2003prb,Ishikawa2004prl,Vogt2003prv,saitoh97} for the former and latter one, respectively. A coexistence of Co ions in the excited (IS or HS) and ground (LS) states in a lattice is expected to cause a sizable disproportionation of Co-O bond lengths. However, this has never been observed despite the effort to do so.

The excitonic scenario is based on the observation that not only
the spin but also the multiplet flavors (LS, IS, and HS)
undergo nearest-neighbor (nn) exchange via the superexchange mechanism.
Inter-atomic exchange processes such as $|{\rm LS,IS}\rangle\leftrightarrow|{\rm IS,LS}\rangle$, see Fig.~1a,
$|{\rm LS,HS}\rangle\leftrightarrow|{\rm IS,IS}\rangle$, or $|{\rm IS,HS}\rangle\leftrightarrow|{\rm HS,IS}\rangle$
turn out to have sizable amplitudes on nn bonds.
At low temperatures, where the state of the system is dominated by atomic LS states,
only the first process is relevant. It gives rise to propagation of IS excitations (excitons) 
on the LS background. As usual in periodic systems, the elementary IS excitations
have the plane-wave form with the energy dependent on the quasi-momentum \textit{\textbf{q}}, see
Fig.~1b. 

The IS excitons come in two orbital symmetries (irreducible representations):
$^3T_{1g}$ ($d_{xy}\otimes d_{x^2-y^2}$, $d_{zx}\otimes d_{z^2-x^2}$, and $d_{yz}\otimes d_{y^2-z^2}$) and 
$^3T_{2g}$ ($d_{xy}\otimes d_{z^2}$, $d_{zx}\otimes d_{y^2}$, and $d_{yz}\otimes d_{x^2}$).
Due to their geometry,
the $^3T_{1g}$ excitons have lower on-site energies (stronger bonding) and larger mobility, concentrated
to their respective planes.
The HS excitations behave differently. 
The nn HS-LS exchange is a fourth-order process in electron hopping and thus has a substantially smaller amplitude than the second-order IS-LS exchange. 
The HS excitation can be approximately treated as an immobile bound pair (bi-exciton) of two IS excitons with different orbital flavors.

Existence of dispersive low-energy excitations has profound consequences. Their 
thermal population does not lead to a static distribution of excited atomic states
and thus does not induce lattice distortions.
When the excitation gap is closed, e.g., by application of strong magnetic field~\cite{Ikeda2016prb}, the excitations with \textit{\textbf{q}}-vector of the band minimum 
form a condensate. 
Recent LDA+U calculations~\cite{Afonso2017prb} find LaCoO$_3$ to be close to the condensation instability. The metamagnetic transition observed in
high fields~\cite{Ikeda2016prb} has the temperature dependence consistent 
with exciton condensation, but not with HS-LS spin-state order~\cite{Sotnikov2016sr}.
The properties of the low-temperature phase of related
(Pr$_{1-y}$R$_y$)$_x$Ca$_{1-x}$CoO$_3$ have been consistently
explained by exciton condensation~\cite{Kunes2014prb_b,Yamaguchi2017jpsj}.
Despite this indirect evidence an unambiguous proof of the excitonic physics in LaCoO$_3$ has been missing. Ultimately, this can be provided by direct observation of the IS dispersion.
In this paper, we present its first observation using Co $L_3$ RIXS technique. Theoretical calculations including the IS states as mobile excitons and immobile HS states support the experimental observation.

\section{method}

\subsection{Experiment}

RIXS has become a powerful tool to study low-energy excitations in transition metal oxides in the last decade~\cite{Ament2011rmp}.
The Co $L_3$-edge RIXS ($2p_{3/2}\rightarrow 3d\rightarrow 2p_{3/2}$) provides sufficient energy resolution to distinguish different spin states~\cite{Tomiyasu2017}. The RIXS amplitude for the IS excitations is sufficient to enable observation of their dispersion, although the X-ray wave-length at the Co $L_3$ edge in LaCoO$_3$ ($\approx15.9$~\AA) restricts the accessible momentum transfer.

\begin{figure}[tbp]
    \includegraphics[width=0.97\columnwidth]{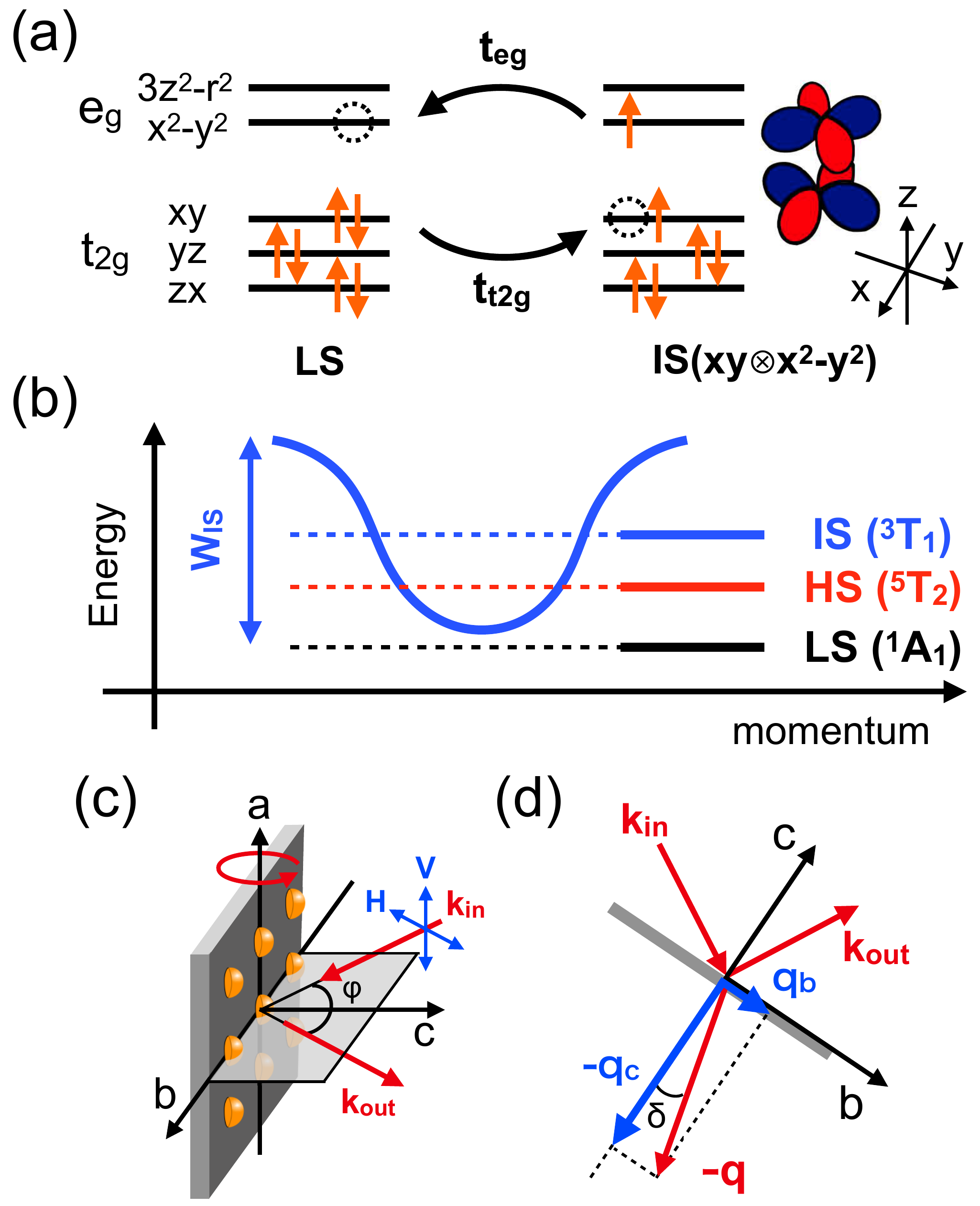}\hspace{0pt}
	\caption{(a) A cartoon view of the nn hopping process 
    with the dominant contribution to the IS propagation,
    and the orbital structure of the $^3T_{1g}$ excitation.
    (b) Sketch of the atomic-level energies together
    with the dispersion of the IS ($^3T_{1g}$) state
	on the LS background in the lattice.
	(c) The experimental geometry and
       the definition of the scattering angle $\varphi$.
	  The sample can be rotated around the $a$ axis.
	  The half-spheres represent Co atoms. 
	(d) Determination of the momentum transfer
      $\textit{\textbf{q}}=\textit{\textbf{k}}_{\rm out}$-$\textit{\textbf{k}}_{\rm in}$.
	The component $q_b$ due to the offset $\delta$ is negligibly small ($|q_b|<0.03\pi$).
	}
	\label{fig1}
\end{figure}

The LaCoO$_3$ single crystal was grown by the optical floating zone method~\cite{Tomiyasu2017}.
The RIXS measurements were performed at the BL05 A1 in Taiwan Light Source (TLS),
with linearly polarized X-rays, either vertical ($V$) or horizontal ($H$) to the scattering plane.
The overall energy resolution at the Co $L_3$ edge ($\sim$780~eV) was 90~meV~\cite{Lai2014jsr}.
The experimental setup is illustrated in Figs.~\ref{fig1}(c) and \ref{fig1}(d).
The sample normal was aligned to the $c$-axis in the (pseudo) cubic axis
with a lattice constant $a_{\rm cub}\approx$~3.83~\AA.
The measurements were carried out in the $bc$ scattering plane by rotating the sample along the $a$-axis.
We define the momentum transfer $\textit{\textbf{q}}=(0, q_b, q_c)/a_{\rm cub}$ as 
the projection of the transferred momentum $\textit{\textbf{q}}$ onto the $b$ and $c$ axes,
and $a_{\rm cub}$ is omitted from now on, for simplicity.
The scattering angles $\varphi$ of 148$^\circ$, 120$^\circ$, 90$^\circ$, and 40$^\circ$
correspond to 
\textit{\textbf{q}}= ($0, q_b, 0.26\pi$),
($0, q_b, 0.48\pi$), ($0, q_b, 0.68\pi$) and ($0, q_b, 0.90\pi$), respectively.
We set a small offset $\delta$ to avoid a strong signal due to reflection, see Fig.~\ref{fig1}(d).
It implies a small $q_b$ projection value~($|q_b|< 0.03\pi $),
that is negligible with the present energy resolution.
The sample temperature was 20~K,~i.e., well below the spin-crossover temperature.
Details of the sample preparation and our data analysis can be found in~Supplementary~Material (SM)~\cite{sm}.

\subsection{Theory}

\begin{figure}[t]
	\includegraphics[width=0.85\columnwidth]{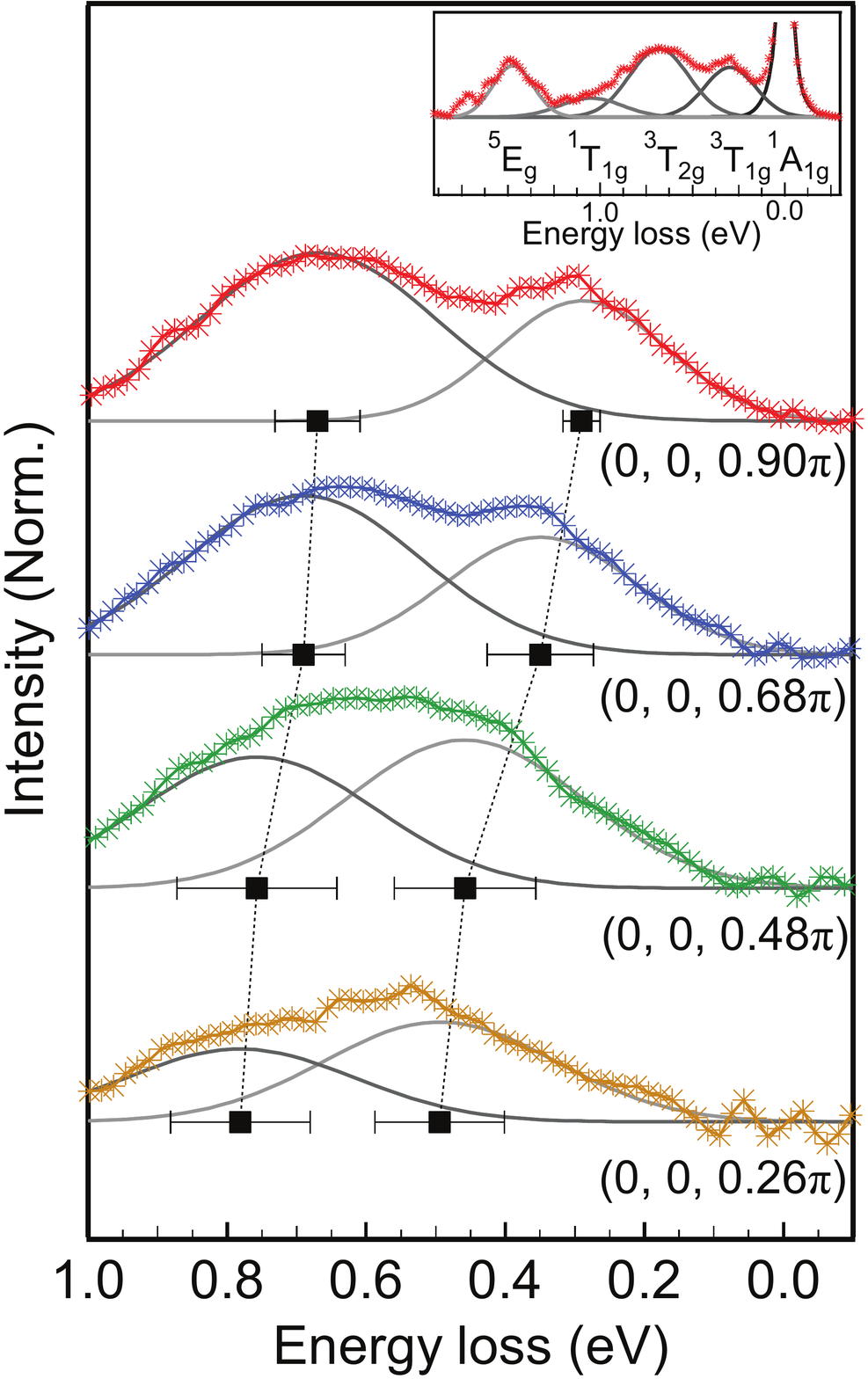}\hspace{0pt}
	\caption{Experimental spectra for different momentum transfers ($0, 0, q_c$).
	The elastic peak is subtracted and each spectrum is fitted by two Gaussian functions with 250 meV of FWHM,
	with centers indicated by squares with the error bar.  
	The inset shows the spectrum at \textit{\textbf{q}}=($0, 0, 0.90\pi$) in a wide energy range
	together with the term symbols~(see discussion in text).}
	 \label{fig2}
\end{figure}

\begin{figure}[t]
	\includegraphics[width=1.05\columnwidth]{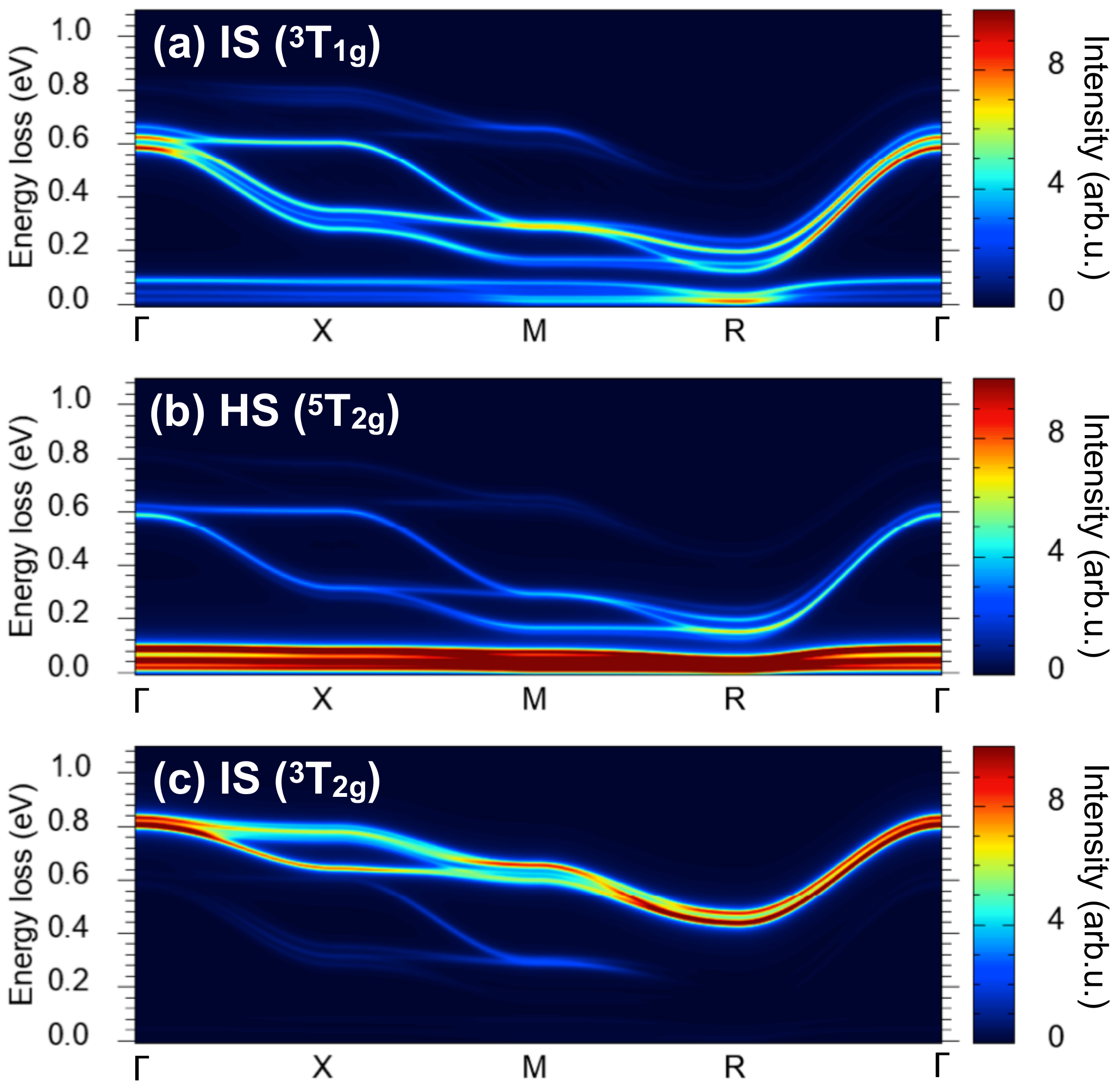}
	\caption{The calculated densities of particle-hole excitations
    resolved into contributions of different atomic multiplets:
    (a) IS ($^3T_{1g}$), (b) HS ($^5T_{2g}$) and (c) IS ($^3T_{2g})$.
    The spectra were artificially broadened with a Lorentzian of 10~meV width.
    The interaction parameters $U$=2.1~eV and $J=0.66$~eV,
    and the SOC amplitude $\zeta_d$=56~meV
    were used in
    the effective Hubbard 'Co $d$-only' model. 
	}
	\label{fig3}
\end{figure}  


\begin{figure}[t]
	\includegraphics[width=1.02\columnwidth]{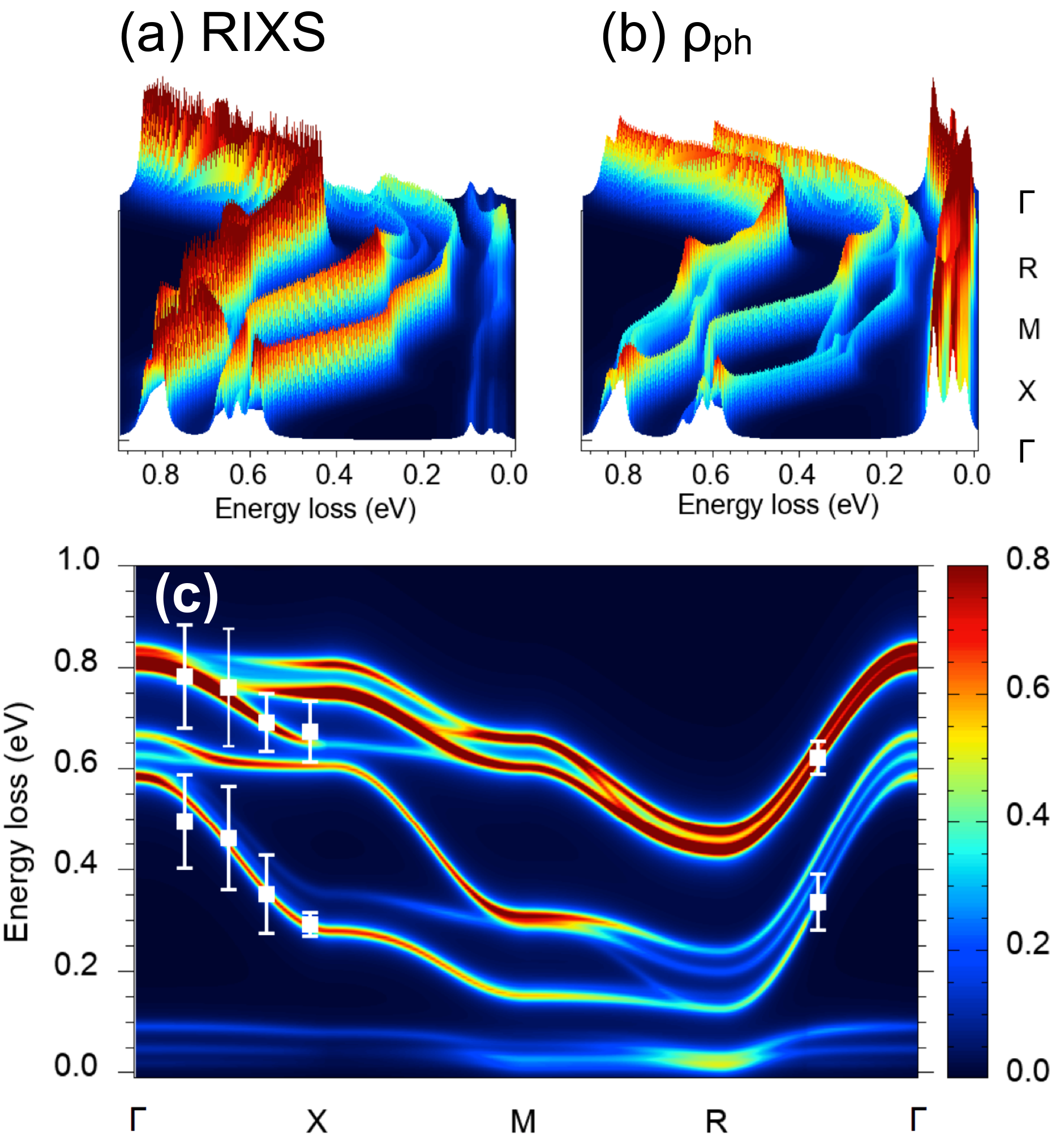}
	\caption{The calculated RIXS intensities (a,c) along the 
    high-symmetry directions in the pseudo-cubic BZ 
    compared with experimental data from Fig.~\ref{fig2}. The
    additional point $\textit{\textbf{q}}$=$(0.52\pi, 0.52\pi, 0.52\pi)$ was measured 
    with $H$ polarization~\cite{sm}.
    The IS ($^3T_{1g}$) excitations show the dispersion feature from 10 to 600~meV.
    The $^3T_{2g}$ IS excitations are located at higher energies.
    The flat bands located below 100~meV correspond to the spin-orbit split HS multiplet.
    (b) 3D plot showing the calculated total density of the particle-hole excitations $\rho_{\rm ph}$.
     Comparison to the RIXS spectra reveals the suppression of the HS intensity by
     the transition matrix elements.
	}
	\label{fig4}
\end{figure}  
 
Theoretical calculation of RIXS spectra is a complicated task.
We adopt the formulation by Haverkort~\cite{Haverkort2010prl}, which factorizes the RIXS cross-section as
\begin{eqnarray}
  \frac{\delta^2\sigma}{\delta\Omega\delta \omega} \propto
     {\rm Im} \sum_{\gamma,\gamma'} R^\dagger_{\gamma'}(\textit{\textbf{q}},\omega_{\rm in})
     G_{\gamma',\gamma}(\textit{\textbf{q}},\omega_{\rm loss})R_{\gamma}(\textit{\textbf{q}},\omega_{\rm in}), \notag
\end{eqnarray}
into X-ray absorption/emission amplitude $R_{\gamma}(\textit{\textbf{q}},\omega_{\rm in})$ and the electron-hole propagator
$G_{\gamma',\gamma}(\textit{\textbf{q}},\omega_{\rm loss})$. While the former determines the intensity (visibility) of different
multiplet excitations $\gamma$ in the RIXS spectra, it is solely the latter one that determines their dispersions.
Here $\omega_{\rm in}$ is the incident photon energy and $\omega_{\rm loss}$ is the energy transfer. $R_{\gamma}(\textit{\textbf{q}},\omega_{\rm in})=\langle\gamma | V_{\boldmath{\epsilon}_{\rm out}}(\omega_{\rm in}+E_{\rm LS}-H+i\Gamma)^{-1}V_{\boldmath{\epsilon}_{\rm in}}|{\rm LS}\rangle$, where the $V_{\boldmath{\epsilon}_{\rm in}} (V_{\boldmath{\epsilon}_{\rm out}})$ operators describe the electron-photon interaction. A sufficiently accurate estimate of the amplitudes $R_{\gamma}(\textit{\textbf{q}},\omega_{\rm in})$ is provided by atomic-model calculation for Co$^{3+}$, which includes  the experimental geometry encoded in $V_{\boldmath{\epsilon}_{\rm in}} (V_{\boldmath{\epsilon}_{\rm out}})$ operators, full-multiplet form of 3$d$-3$d$ and 2$p$-3$d$ Coulomb interaction, the $3d$ crystal field and the spin-orbit coupling (SOC) in the Co 3$d$ shell and the 2$p$ shell. The SOC within the 2$p$ shell $\zeta_{p}$ and the Slater integrals for the 2$p$-3$d$ interaction $F^k$, $G^k$ are calculated within an atomic Hartree-Fock code. Then the $F^k$ and $G^k$ values for the 2$p$-3$d$ interaction are scaled down by the empirical factor 75\% to simulate the effect of intra-atomic configuration interaction from higher basis configurations neglected in the atomic calculation. We note that the amplitude $R_{\gamma}(\textit{\textbf{q}},\omega_{\rm in})$ for the LS ground state is rather insensitive to the choice of the empirical factor.

The particle-hole propagator $G_{\gamma',\gamma}(\textit{\textbf{q}},\omega_{\rm loss})$, which determines dispersion of the excitations, is the key theoretical quantity studied in this work. Its evaluation is, in general, a difficult task that requires approximations. The insulating ground state of LaCoO$_3$~\cite{Chainani92,Arima93,Yamaguchi96}, which can be viewed as a collection
of LS atoms allows us to eliminate the local charge fluctuations and to use a low-energy effective model where only a few atomic multiplets and their nearest-neighbor interactions are retained. The description is further simplified at the low-temperatures ($T\lesssim 20$~K) where thermal excitations can be neglected and generalized spin-wave theory can be used, which describes the excitations as non-interacting bosons propagating on the lattice.

Construction of the model starts with the density-functional calculation for the idealized cubic perovskite structure ($a$=3.8498~\AA) using Wien2k~\cite{wien2k}. Then, an effective Hubbard model spanning the Co 3$d$-like band is obtained with the wien2wannier~\cite{wien2wannier} and wannier90~\cite{wannier90} software. The intra-$d$-shell interaction is parametrized with $U$ and $J$, treated as adjustable parameters of our theory. Next, second order Schrieffer--Wolff transformation~\cite{sw66} into the low-energy effective model, spanning the LS, HS ($^5T_{2g}$), and IS ($^3T_{1g}$ and $^3T_{2g}$) states, is performed.
We include spin-orbit coupling (SOC), but neglect the rhombohedral distortion of the real LaCoO$_3$ structure. This is still a complicated many-body model which involves 34 states per Co site. A substantial simplification is achieved at low temperatures when the system is in its ground state, which can be well approximated as a product of LS states on every site. The remaining 33 states can be viewed as bosonic excitations, which can propagate between sites with possible change of flavor. Their dispersion can be obtained with the generalized spin wave approach~\cite{Sommer01}, which leads to effective Hamiltonian
\begin{eqnarray}\label{ham_BH}
    H_{\rm eff}&=&H_{0}+H_{\rm int}, \notag \\
    { H}_{0} &=&
	\sum_{ij,\,\gamma\gamma'}(h_{\gamma\gamma'}^{ij}
	{d}^\dag_{i\gamma} {d}_{j\gamma'}
	+ \Delta_{\gamma\gamma'}^{ij}
	{d}^\dag_{i\gamma} {d}^\dag_{j\gamma'} 
    +{\rm{H.c.}}). 
\end{eqnarray}
Here, we view the product of atomic LS-like ground states~\footnote{This is not the exact LS state due to mixing with other atomic states by spin-orbit couping.} as the vacuum
and the other 33 states of the HS and IS states as different bosonic excitations $\gamma$ characterized by the corresponding creation (annihilation) operators $ {d}^{\dag}_{i\gamma}$ ($ {d}_{i\gamma}$) on the lattice site $i$. The term with the amplitude $h$ corresponds to the renormalized on-site energies of bosons (for $i=j$), and their hopping amplitudes on the LS background (for $i\neq j$). The term with the amplitude $\Delta$ describes the non-local pair-creation/annihilation processes. The interaction term $H_{\rm int}$ including third and fourth order terms in $d$ ($d^{\dag}$) can be neglected in low temperatures of LaCoO$_3$ since the density of these excitations is negligible in the LS insulating ground state.
Finally, we treat (\ref{ham_BH}) numerical Bogoliubuv transformation.
Details of the model construction can be found in SM~\cite{sm} and Ref.~\onlinecite{Afonso2018}.

\section{Results and Discussion}

Fig.~\ref{fig2} shows experimental RIXS spectra along the path from $\Gamma$($0, 0, 0$)  to $X$($0, 0, \pi$)
recorded at 20~K, well below the spin-crossover temperature.
The spectra were normalized to the fluorescence that was subsequently subtracted, see SM for details~\cite{sm}. The inset shows the spectrum at \textit{\textbf{q}}=($0, 0, 0.90\pi$) in a wide energy window. It can be decomposed into five Gaussian contributions with the full width at half maximum (FWHM) of 250 meV, accounting for the instrumental resolution ($\Delta E=90$~meV), the spin-orbit spitting of the multiplets, and possible vibrational effect~\cite{lee14}. The four peaks at around 0.4, 0.7, 1.2, and 1.6 eV, are attributed to the excitations from LS ($^1A_{1g}$) ground state to IS ($^3T_{1g}$), IS ($^3T_{2g}$), LS ($^1T_{1g}$), and HS ($^5E_{g}$) states, respectively~\cite{Tomiyasu2017}. We point out that the lowest HS ($^5T_{2g}$) state, located below 100 meV~\cite{Haverkort2006prl}, has a negligible RIXS intensity
in the LS ground state~\cite{Tomiyasu2017,Groot98} and thus is not visible at low temperature. The IS $^3T_{1g}$ peak exhibits a clear $\textit{\textbf{q}}$-dependent shift from 490 to 290 meV in the interval from \textit{\textbf{q}}=($0, 0, 0.26\pi$) to ($0, 0, 0.90\pi$). The $\textit{\textbf{q}}$-dependence of the IS $^3 T_{2g}$ peak at around 0.7~eV is much less pronounced.

The excitation spectra in Fig.~\ref{fig3} were obtained for $U$=2.1~eV and $J$=0.66~eV. These values lead to the best match with the present RIXS data at 20~K in addition to fulfilling following experimental constraints:~(1) the ground state at 20~K is the insulating singlet state~\cite{Chainani92,Arima93,Yamaguchi96} that does not exhibit any symmetry breaking;~(2)~the energy of the lowest HS excitation falls into 10-20~meV range according to inelastic neutron scattering studies~\cite{Podlesnyak2006prl,Noguchi2002prb}. As shown in Appendix~\ref{App.A}, 
the calculated spectra are quite sensitive to the variation of $U$ and $J$, which
has a complex effect of changing simultaneously the positions of band centers and bandwidths of all excitations. Therefore the existence of parameters matching all experimental constraints is more than trivial fitting. When comparing with interaction
parameters used in other studies one has to keep in mind that
$U$ and $J$ are strongly basis dependent. Therefore the present values can be compared to studies based on 3$d$-only models~\cite{karolak15}, but not to the values used in  LDA+$U$ or models including the O $2p$ states explicitly.


Fig.~\ref{fig3} shows the contributions of different atomic multiplets 
to the calculated dispersion of elementary excitations.
The sizable dispersion of the IS $^3T_{1g}$ branch, describing a propagation
of a single IS $^3T_{1g}$ state on the LS background, originates
from processes such as the one depicted in Fig.~\ref{fig1}(a). The band minimum
at the $R$ point is a simple consequence of the electron nn hopping amplitudes
$t_{e_g}$ and $t_{t_{2g}}$ [see Fig.~\ref{fig1}(a)] having the same sign~\cite{Afonso2017prb},
a general feature of the perovskite structure. The enhanced low-energy
IS $^3T_{1g}$ intensity around the $R$ point is partly due to weak nn pair creation/anihilation processes $|{\rm LS,LS}\rangle\leftrightarrow|{\rm IS,IS}\rangle$~\cite{sm,Afonso2018}.
The SOC induces a splitting in the IS $^3T_{1g}$ states, which is clearly seen along the $R$-$\Gamma$ direction, see Fig.~\ref{fig3}a and Fig.~\ref{fig4}c.

The HS excitations, within the
present approximation, have no hopping on the LS background.
As a result, they form almost flat bands with the centers at approximately 20, 45 and 90~meV, split and partially mixed with IS by the SOC.
These energies are consistent with other studies~\cite{Noguchi2002prb,Tomiyasu2017,Podlesnyak2006prl}.

Inclusion of the transition amplitudes
$R_{\gamma}(\textit{\textbf{q}},\omega_{\rm in})$
strongly suppresses the contribution of the HS ($^5T_{2g}$) states to the RIXS spectra,
see Fig.~\ref{fig4}(a) and \ref{fig4}(b).
The calculated RIXS intensities along high-symmetry lines in the cubic
Brillouin zone (BZ)
together with the experimental $^3T_{1g}$ and $^3T_{2g}$ peak positions
are shown in Fig.~\ref{fig4}(c).
We find a very good match in the experimentally 
accessible part of BZ along the $\Gamma$-$X$ and $\Gamma$-$R$ directions.

The most interesting region 
around $R$ point is out of the experimental reach at Co $L_3$-edge.
Due to the dominant nn character of the exciton hopping, the shape of the dispersion is largely determined by the lattice structure. Knowing the dispersion over a substantial 
part of the bandwidth thus puts the extrapolation on a solid ground.

Without experimental data about the detailed excitation spectrum around the $R$ point
we can speculate about two possible scenarios:
(i) the lowest excitation is dominantly IS, 
implying that exciton condensation would be possible upon closing of the excitation
gap, e.g., by a magnetic field as discussed~\cite{Sotnikov2016sr}
in some interval
of small IS concentrations;
(ii) the lowest excitation is
dominantly HS, but the existence of mobile IS excitations prevents a formation
of the spin-state (HS-LS) order due to $|{\rm HS,LS}\rangle\leftrightarrow| {\rm IS,IS}\rangle$ fluctuations.

\section{Conclusions}
Using Co $L_3$-edge RIXS in LaCoO$_3$ at 20~K we have observed dispersion of the IS ($^3T_{1g}$) excitations with a sizable bandwidth. The experimental data match well the theoretical calculations 
in the experimentally accessible part of the Brillouin zone
and their extrapolation points to an important role of IS excitations for the low-energy physics of the material. Dispersion dominated by the nearest-neighbor processes allows for reliable extrapolation. Improved energy resolution and polarization analysis would reduce reliance on theoretical model and is therefore highly desirable.

The present results show that the question, whether the first atomic excited state is HS or IS, is not correctly posed. While there is little doubt that the lowest atomic excitation is HS, the propensity of IS states to move on the LS lattice changes the game in the extended system.
LaCoO$_3$, therefore, should not be viewed as a static collection of ions in particular atomic states, but rather as a gas of mobile bosonic excitons (IS) above (LS) vacuum. 
The HS states play the role of strongly bound and essentially immobile bi-excitons. 
This picture provides a natural explanation why the spin-state order accompanied by Co-O bond-length disproportionation is not observed in LaCoO$_3$ despite the low-energy of HS excitations.
It further suggests LaCoO$_3$ and its analogs as potential materials for realization of excitonic magnetism~\cite{Khaliullin13}.\\

\begin{acknowledgments}
The authors thank J. Fern\'andez Afonso for fruitful discussions,
You-Hui Liang for supporting the XRD measurement,
and Ties Haarman for providing the fitting codes.
A.H., A.So. and J.K. are supported by the European Research Council (ERC)
under the European Union’s Horizon 2020 research and innovation programme (grant agreement No.~646807-EXMAG).
D.J.H. was supported by the Ministry of Science and Technology of Taiwan under
Grant No.~103-2112-M-213-008-MY3.
The experiments were supported by ERC advanced grant (grant agreement No.~340279-XRAYonACTIVE).
K.T. is financially supported by the MEXT and JSPS KAKENHI (JP17H06137, JP15H03692).
We thank technical staff of Taiwan Light Source for help with RIXS measurements.
The calculations were performed on Vienna Scientific Cluster (VSC).\\

\end{acknowledgments}

\begin{appendix}
\section{Parameter dependence of particle-hole excitations}\label{App.A}

\begin{figure}[t]
	\includegraphics[width=1.02\columnwidth]{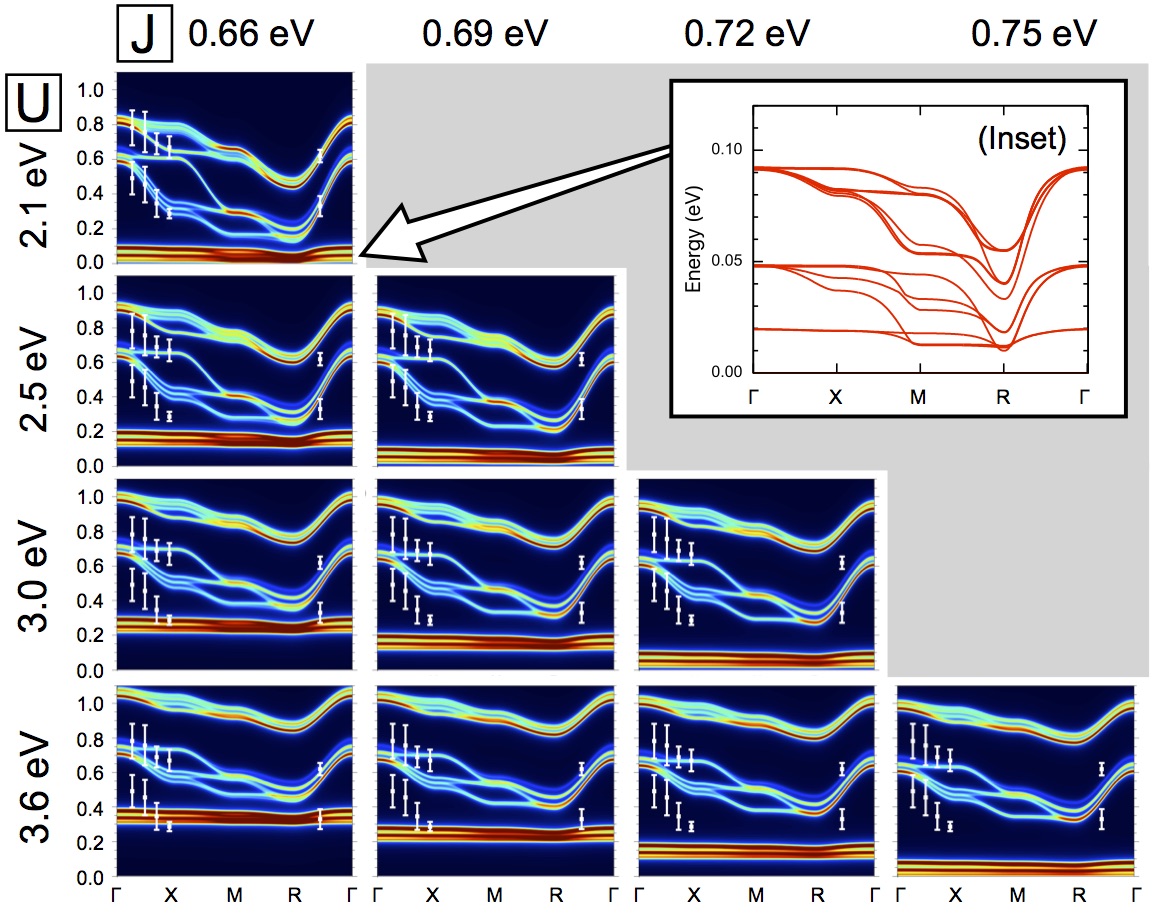}
	\caption{ The calculated total densities of particle-hole excitations $\rho_{\rm ph}$ for different $U$ (row) and $J$ (column) values, together with the RIXS experimental data (white points with error bars) from Fig~2. The inset shows the low-energy region for ($U$, $J$)=(2.1~eV, 0.66~eV).}
	\label{fig_scan}
\end{figure}

Figure~\ref{fig_scan} shows the theoretically calculated spectra of particle-hole excitations $\rho_{\rm ph}$ in LaCoO$_3$ for different $U$ (row) and $J$ (column) values, together with the present RIXS experimental data shown in the main part (see Figs.~\ref{fig2} and \ref{fig4}). For the parameters in the upper-right triangle (gray background) or at smaller $U$ and $J$ values, the LS state is not a stable ground state, contradicting the experimental observations at $T\lesssim 20$~K~\cite{Asai1998jpsj,Stolen1997prb,Haverkort2006prl}. The excitation energy of the lowest high-spin state ($\sim$10--20 meV) is in agreement with other studies~\cite{Podlesnyak2006prl,Noguchi2002prb,Tomiyasu2017} only along the diagonal panels in Fig.~\ref{fig_scan}. For ($U$, $J$)=(2.1~eV, 0.66~eV), the best agreement with the RIXS experimental data is found.
The inset shows the low-energy region for the chosen regime. 

\end{appendix}


\bibliography{rixs}

\end{document}